# A simple Bayesian procedure for forecasting the outcomes of the UEFA Champions League matches

Jean-Louis Foulley[1]

**Abstract**:
This article presents a Bayesian implementation of a cumulative probit model to forecast the outcomes of the UEFA Champions League matches. The argument of the normal CDF involves a cut-off point, a home vs away playing effect and the difference in strength of the two competing teams. Team strength is assumed to follow a Gaussian distribution the expectation of which is expressed as a linear regression on an external rating of the team from eg. the UEFA Club Ranking (UEFACR) or the Football Club World Ranking (FCWR). Priors on these parameters are updated at the beginning of each season from their posterior distributions obtained at the end of the previous one. This allows making predictions of match results for each phase of the competition: group stage and knock-out. An application is presented for the 2013-2014 season. Adjustment based on the FCWR performs better than on UEFACR. Overall, using the former provides a net improvement of 24% and 23% in accuracy and Brier's score over the control (zero prior expected difference between teams). A rating and ranking list of teams on their performance at this tournament and possibilities to include extra sources of information (expertise) into the model are also discussed.

---

[1] Affiliated to Institut de Mathématiques et de Modélisation de Montpellier (3M) Université de Montpellier II 34095 Montpellier Cedex 05 and E-mail : Jean-Louis.Foulley@math.cnrs.fr

# 1. Introduction

There is a growing interest in statistics on sport and competitions from both a theoretical and an applied point of view due in particular to the availability of a large amount of data, the development of betting internet sites and the large media coverage of sporting events. Major competitions such as the UEFA Champions League and FIFA World Cup offer great opportunities to test and implement various models for analyzing the data generated by these competitions. These kinds of competitions are especially appealing to statisticians as they involve two different steps: i) a round-robin tournament where each contestant meets all the other ones within each group, followed for the top ranked teams by ii) a knock-out stage in which only the winners of each stage (round of 16, quarter and semi-finals) play the next stage up to the final. These two steps raise some difficulties for the statistician especially due to the differences of group levels of teams drawn for the mini-championship that can affect the choice of teams qualified for and playing in the next round. Another key aspect consists of fitting models to data either for rating & ranking teams or for forecasting outcomes of forthcoming matches. Although these two goals are not disconnected, they require different procedures and criteria to assess their efficiency. Knowing that, Section 2 presents a brief description of the competition and its different stages. Statistical methods are expounded in Section 3 with a focus on the cumulated probit model, its Bayesian implementation and its use in forecasting match results. An application to the 2013-2014 season is displayed in Section 4 to illustrate these procedures. Finally, Section 5 summarizes this study and discusses several alternatives and possible improvements of the model.

## 2. The tournament

The tournament per se begins with a double round group stage of 32 teams distributed into eight groups: A, B, C, D, E, F, G and H. The eight groups result from a draw among the 32 teams allocated into four "pots" based on their UEFA club coefficients (Table 1). Among the 32 participants, 22 were automatically qualified and the ten remaining teams were selected through two qualification streams for national league champions and non champions. In the case studied here (2013-14 season), the 22 teams were: the title holder of the previous season (Bayern Munich for 2013-2014), the top three clubs of England (GBR), Spain (ESP) and Germany (DEU) championships, the top two of Italy (ITA), Portugal (PRT) and France (FRA) and the champions of Russia (RUS), Ukraine (UKR), Netherlands (NLD), Turkey (TUR), Denmark (DNK) and Greece (GRC).

Table 1: Distribution of the 32 qualified teams for UEFA 2013-14 into the four pots and their UEFA coefficient

| Pot 1 | Pot 2 | Pot 3 | Pot 4 |
|---|---|---|---|
| Bayern Munich*-DEU 146.922 | Atlético Madrid-ESP 99.605 | *Zenith Petersburg*-RUS 70.766 | FC Copenhagen-DNK 47.140 |
| FC Barcelona-ESP 157.605 | Shakhtar Donetsk-UKR 94.951 | Manchester City-GBR 70.592 | SSC Napoli-ITA 46.829 |
| Chelsea -GBR 137.592 | *AC Milano-ITA 93.829* | Ajax Amsterdam-NLD 64.945 | RSC Anderlecht-BEL 44.880 |
| Real de Madrid -ESP 136.605) | *Schalke 04-DEU 84.922* | Borussia Dortmund-DEU 61.922 | *Celtic Glasgow-SCO 37.538* |
| Manchester United-GBR 130.592 | O Marseille FRA 78.800 | *FC Basel-CHE 59.785* | *Steaua Bucarest-ROU 35.604* |
| *Arsenal-GBR 113.592* | CSKA Moscow -RUS 77.766 | Olympiacos-GRC 57.800 | Viktoria Plzen-*CZE 28.745* |
| *FC Porto-PRT 104.833)* | Paris Saint Germain-FRA 71.800 | Galatasaray SK-TUR 54.400 | *Real Sociedad-ESP 17.605* |
| Benfica Lisbon-PRT 102.833 | Juventus-ITA 70.829 | Bayer Leverkusen –DEU 53.922 | *Austria Wien-AUT 16.575* |

Team underlined: National League Champion; Team in italics: qualified through the play-off rounds
*Title holder of the UEFA Champions League 2012-13

The group stage played in autumn consists of a mini championship of twelve home and away matches between the four teams of the same group. The winning team and the second ranked of each group progress to the next knock out stage made up of home and away matches among the winning team of one group and the second of another group after a random draw held in December excluding teams from the same association country for the round of 16. This exclusion rule does not apply later on (eighth, quarter and semi finals), the final being played in a single match on neutral ground. Points are based on the following scoring system: 3, 1, and 0 for a win, draw and loss respectively with a preference for the goals scored at the opponent's stadium in case of a tied aggregate score.

### 3. Statistical methods

*3.1 A benchmark model*

Outcomes of matches under the format of Win Draw and Loss can be predicted either directly or indirectly via the number of goals scored by the two teams. As there is little practical difference between these two approaches (Goddard, 2005), we use for the sake of simplicity the former benchmark model under its latent ordered probit form known in sports statistics, as the Glenn and David (1960) model.

Revised 12-02-15

Let us briefly recall the reader the structure of this model. The random variable pertaining to the outcome $X_{ij}$ of the match $m(ij)$ between home $i$ and away $j$ teams is expressed via the cumulated density function of an associated latent variable $Z_{ij}$ as follows (Agresti, 1992):

$$\pi_{ij,1} = \Pr(X_{ij} = 1) = \Pr(Z_{ij} > \delta) \quad (1a)$$

$$\pi_{ij,2} = \Pr(X_{ij} = 2) = \Pr(-\delta \leq Z_{ij} \leq \delta) \quad (1b)$$

$$\pi_{ij,3} = \Pr(X_{ij} = 3) = \Pr(Z_{ij} < -\delta) \quad (1c)$$

with $\delta$ being the threshold or cut-off point on the underlying scale and 1,2,3 standing for win, draw and loss respectively.

Assuming that $Z_{ij}$ has a Gaussian distribution with mean $\mu_{ij}$ and unit standard deviation, the cumulative probit model consists of expressing $\mu_{ij}$ as the sum of the difference $\Delta s_{ij} = s_i - s_j$ in strength between teams $i$ and $j$ and a home vs away playing effect $h_{ij}$ usually taken as a constant $h$ so that:

$$\mu_{ij} = \Delta s_{ij} + h \quad (2)$$

$$\pi_{ij,1} = L(\delta - \Delta s_{ij} - h) \quad (3a)$$

$$\pi_{ij,3} = L(\delta + \Delta s_{ij} + h) \quad (3b)$$

$$\pi_{ij,2} = 1 - \pi_{ij,1} - \pi_{ij,3} \quad (3c)$$

where $L(x) = \Pr(X > x) = 1 - F(x)$ is the survival function equal to 1 minus the CDF $F(x)$, here the standard normal. The higher the difference between the strengths of the two teams $i$ and j, the higher is the probability of win by team $i$ against $j$ which makes sense; the same reasoning applies to the well known home vs away playing advantage.

### 3.2 Bayesian implementation

Now, there are several ways to implement statistical methods for making inference about the parameters of this basic model. Classical methods consider the team strength and home effects as fixed and make inference about them and other covariates effects, using maximum likelihood procedures: see eg. the general review by Cattelan (2012) in the context of the Bradley-Terry model. For others, model fitting to data is accomplished within a Bayesian framework (eg. Glickman, 1999) and this is the way chosen here.

The first stage of the Bayesian hierarchical model reduces to a generalized Bernoulli or categorical distribution Cat(.) with probability parameters of the three possible outcomes described in (3abc)

1) $$X_{ij} | \boldsymbol{\theta} \sim Cat(\boldsymbol{\Pi}_{ij}) \quad (4)$$

where $\boldsymbol{\theta}$ refers to all the model parameters and $\boldsymbol{\Pi}_{ij} = (\pi_{ij,k})$ for $k = 1, 2, 3$.

A the second stage, we have to specify the distributions of parameters involved in $\mathbf{\Pi}_{ij}$ viz. team strength, cut-off value and home effect. This is done as follows:

2) $\quad s_i | \eta_i, \sigma^2 \sim_{id} \mathcal{N}\left(\eta_i, \sigma_s^2\right)$ (5)

where the team strength $s_i$ is assumed normally distributed with mean $\eta_i$ and variance $\sigma^2$.

Similarly, Gaussian distributions are taken for $\delta$ and $h$:

$$\delta \sim \mathcal{N}\left(\delta_0, \sigma_\delta^2\right) \quad (6)$$
$$h \sim \mathcal{N}\left(h_0, \sigma_h^2\right) \quad (7)$$

with means and variances calibrated as explained later on.
At the third stage, prior distributions are:

3) $\quad \eta_i = \beta \tilde{x}_i$ with $\beta \sim \mathcal{N}\left(\beta_0, \sigma_\beta^2\right)$ (8)

where $\tilde{x}_i = (x_i - \bar{x})/\hat{\sigma}$ is a centered standardized reference value for team $i$ and $\beta$ a regression coefficient.

As far as team strength variability is concerned, several choices may be envisioned. A lognormal distribution on the standard deviation was chosen for practical reasons as advocated by Barnard et al. (2000) and Foulley and Jaffrezic (2010):

$$\gamma_s = \log(\sigma_s) \sim \mathcal{N}\left(\gamma_0, \sigma_\gamma^2\right) \quad (9)$$

At the beginning of a new season, the latest evaluations of the 32 teams entering the group stage are incorporated into (5) from exogenous rating systems: UEFA Club Ranking (UEFACR) or Football Club World Ranking (FCWR).

Similarly, the parameters of the prior distributions of $\delta$, $h$, $\beta$ and $\gamma_s$ in (6,7,8,9) are updated from their posterior distributions calculated at the end of the previous season. This can be done recursively so that past information on the matches played in all the preceding seasons is automatically taken into account in the model.

Posterior inference of the parameters is based on a Gibbs sampling algorithm. This can be easily carried out using the Winbugs/Openbugs software (Lunn et al., 2013).

### *3.3 Prediction and its efficiency*

Prediction of outcomes of forthcoming matches is based on the marginal posterior predictive distribution of $X_{ij}^{new} = k | \mathbf{y}^{av}$ given available information $\mathbf{y}^{av}$ up to the time of the match $m(ij)$. $\mathbf{y}^{av}$ includes information from the previous season for the matches of the group stage and results from the group stage and additional information from the previous rounds of the knock out for this second phase (eg. round of 16 used to predict outcomes of the quarter finals).

Efficiency of prediction is assessed by two criteria: the Brier score and accuracy.

The Brier score (1950) for any match $m$ is defined as the squared difference between the probabilities of the different forecast outcomes of the match $P_{m,k}(\theta)$ and the actual one when observed $O_{m,k}$:

$$B_m = \sum_{k=1}^{K=3}\left[P_{m,k}(\boldsymbol{\theta}) - O_{m,k}\right]^2 \qquad (10)$$

Here $P_{m,k}(\boldsymbol{\theta}) = \Pr(X_m^{new} = k \mid \boldsymbol{\theta})$ and $O_{m,k} = I\left(X_m^{obs} = k\right)$ with $I(.)$ being the indicator variable. Notice that $B_m$ varies in the range 0 (exact forecast) to 2 (false forecast with a probability of one). For a set of $M$ matches, we just take: $B = \left(\sum_{m=1}^{M} B_m\right)/M$. As $B_m$ is a function of the $\boldsymbol{\theta}$ parameters of the model, there are two ways of estimating it: either as a plug-in estimator replacing $\boldsymbol{\theta}$ by its posterior mean $\bar{\boldsymbol{\theta}} = E(\boldsymbol{\theta} \mid \mathbf{y}^{av})$ or, as here, by its posterior predictive expectation: $\bar{B}_m = E(B_m \mid \mathbf{y}^{av})$.

In this formulation (10), the Brier score is derived from a quadratic discrepancy function of observed data and parameters as defined in Gelman et al. (2004, chapter 6) and which can be viewed as an analog of the deviance function used in model comparison. As noted by Gelman et al. (2004) and Plummer (2008), the expected form has some advantages over the plug-in; it is insensitive to reparameterization and takes the precision of parameters into account.

As some people might be not familiar with the scale of Brier's score, we also present a more accessible criterion of efficiency namely "Accuracy". Accuracy (A) or Exact Forecasting Rate (EFR) is defined here as the expected percent of correctly forecasted outcomes of matches. For a given match, the accuracy of the forecast $A_m$ is defined as

$$A_m = \Pr(X_m^{new} = X_m^{obs} \mid \boldsymbol{\theta}) \qquad (11)$$

and is estimated by $\bar{A}_m = E(A_m \mid \mathbf{y}^{av})$ which is the posterior predictive probability that the forecast outcome of the match (Win or Draw or Loss) is the actual one. As previously regarding the Brier score, $A$ is taken as the arithmetic mean $A = \left(\sum_{m=1}^{M} A_m\right)/M$ for a set of $M$ matches. Notice also that A can be viewed as particular case of the general criteria proposed by Laud and Ibrahim (1995) to assess model efficiency from an expected distance between observed and predictive distributions (0-1 loss).

## 4. Application to the 2013-14 Champions league

### *4.1 Data and model implementation*

The composition of the four pots giving rise to the draw of August 29, 2013 for the 2013-14 round-robin tournament is shown in Table 1 along with its result allocating the 32 qualified teams to the eight groups A to G and H (Table 2).

Table2: List of the 32 teams qualified for the group stage of the 2013-14 UEFA Champions league

| No | Group | Team | Abbre | UEFACR | FCWR |
|---|---|---|---|---|---|
| 1 | A | Manchester United (GBR) | MUD | 130.592 | 11537 |
| 2 | | Shakhtar Donetsk (UKR) | SHA | 94.951 | 7142 |
| 3 | | Bayer Leverkusen (DEU) | BLE | 53.922 | 7959 |
| 4 | | *Real Sociedad (ESP)* | RSO | 17.605 | 7091 |
| | | Overall | | *-0.08* | *0.04* |
| 5 | B | Real de Madrid (ESP) | RMA | 136.605 | 14685 |
| 6 | | Juventus (ITA) | JUV | 70.829 | 10046 |
| 7 | | Galatasaray (TUR) | GAL | 54.4 | 6392 |
| 8 | | FC Copenhagen(DNK) | KOB | 47.14 | 2531 |
| | | Overall | | *-0.00* | *0.04* |
| 9 | C | Benfica Lisbon(PRT) | BEN | 102.833 | 10362 |
| 10 | | Paris Saint Germain (FRA) | PSG | 71.8 | 10042 |
| 11 | | Olympiacos (GRC) | OLY | 57.8 | 6588 |
| 12 | | RSC Anderlecht (BEL) | AND | 44.88 | 5004 |
| | | Overall | | *-0.21* | *-0.08* |
| 13 | D | Bayern Munich (DEU) | BAY | 146.922 | 16927 |
| 14 | | CSKA Moscow (RUS) | CSK | 77.766 | 4747 |
| 15 | | Manchester City (GBR) | MCI | 70.592 | 9145 |
| 16 | | *Viktoria Plzen (CZE)* | PLZ | 28.745 | 5655 |
| | | Overall | | *0.10* | *0.24* |

Teams are sorted in each group according to increasing order of draw in the 4 pots
Overall: UEFACR & FCWR group estimate in standard deviation units with SE=0.562 and 0. 564 and F group statistics=0.07 and 0.05 for UEFACR and FCWR respectively

The draw looks fair as it did not generate much difference among the groups as shown by the F statistics of the ANOVA: 0.07 on the UEFA coefficient scale. On this scale, the top group is H (+0.30±0.562 in standard deviation unit) mainly due to the presence of Barcelona ranked second in the pot 1 and the bottom one is C (-0.21±0.562) due to Benfica ranked last in the same pot.

Regarding the reference team value entering as the expectation of the distribution of team strength, two external rating systems were chosen:

    i) the "UEFA Club Ranking" (UEFACR) acting as the official basis for seeding of clubs entering the European competitions (Champions League CL and Europa League EL) , and

    ii) the "Football Club World Ranking" (FCWR) issued by the Institute of Football Club Coaching Statistics (Netherlands) which is a major independent provider of world statistics for football clubs publishing rating and ranking of teams updated weekly.

These two are contrasted with a control situation in which the expectation of team strength is set to zero ($\eta_i = 0$ for any $i$).

Although substantially correlated (r=0.807 with a 95% confidence interval of [0.628; 0.905]), evaluations i) and ii) are not based on the same principles and use historical data on match outcomes differently. The UEFACR takes into account the club performance over five previous UEFA CL an EL competitions. FCWR relies on matches played over the last 52 weeks both at the national and international levels using a complex system of weights.

Table2: List of the 32 teams qualified for the group stage (continued)

| No | Group | Team | Abbrev | UEFACR | FCWR |
|----|-------|------|--------|--------|------|
| 17 | E | Chelsea (GBR) | CHE | 137.592 | 12292 |
| 18 |   | *Schalke 04 (DEU)* | SCH | 84.922 | 7387 |
| 19 |   | *FC Basel (CHE)* | BAL | 59.785 | 5850 |
| 20 |   | *Steaua Bucarest (ROU)* | BUC | 35.604 | 4911 |
|    |   | Overall |     | *0.06* | *-0.19* |
| 21 | F | *Arsenal (GBR)* | ARS | 113.592 | 10257 |
| 22 |   | Olympique Marseille (FRA) | OMA | 78.8 | 5212 |
| 23 |   | Borussia Dortmund (DEU) | DOR | 61.922 | 12110 |
| 24 |   | SSC Napoli (ITA) | NAP | 46.829 | 6436 |
|    |   | Overall |     | *-0.05* | *0.06* |
| 25 | G | *FC Porto (PRT)* | POR | 104.833 | 8078 |
| 26 |   | Atletico Madrid (ESP) | AMA | 99.605 | 13096 |
| 27 |   | *Zenith St Petersburg* (RUS) | ZSP | 70.766 | 6966 |
| 28 |   | *Austria Wien (AUT)* | AWI | 16.575 | 3219 |
|    |   | Overall |     | *-0.12* | *-0.12* |
| 29 | H | FC Barcelona (ESP) | BAR | 157.605 | 14987 |
| 30 |   | *AC Milano (ITA)* | ACM | 93.829 | 7656 |
| 31 |   | Ajax Amsterdam (NLD) | AJX | 64.945 | 5583 |
| 32 |   | *Celtic Glasgow (SCO)* | CEL | 37.538 | 5116 |
|    |   | Overall |     | *0.30* | *0.02* |

To make comparisons between the two systems fair, values of UEFACR and FCWR calculated at the same time ie. at the beginning of the tournament (end of August) were implemented in the model for the two rating systems (Table 2).
This operation can be repeated each new season according to the teams coming in and out of the competition and the updated levels of all the participating teams.
Updates of the prior distributions of cut-off $\delta$, home effect $h$, regression coefficient $\beta$ and precision $\gamma$ were calculated from their posterior distributions obtained at the end of the 2012-13 season. For the sake of simplicity, the same values of parameters were adopted for cut-off and home effects, but different ones have been used for the regression coefficient and the precision as these clearly depend on the type of pre-adjustment (Zero, UEFACR, FCWR) considered (Table 3).

Table 3. Characteristics of the prior distributions used for the parameters of the model according to the type of adjustment

| Parameter | Type of adjustment | | |
|---|---|---|---|
| | Zero | UEFACR | FCWR |
| δ | N(0.335,1/300) | N(0.335,1/300) | N(0.335,1/300) |
| h | N(0.225,1/100) | N(0.225,1/100) | N(0.225,1/100) |
| β | | N(0.250,1/100) | N(0.430,1/120) |
| γ | N(-1.00,1/5.79) | N(-1.13,1/5.00) | N(-2.00,1/2.30) |

N(mean, variance)

## *4.2 Forecasting performance*

Using the prior distributions defined previously from the information available at the end of CL 2012-13 and the external team rating system (UEFACR or FCWR) at the beginning of CL 2013-2014, we are able to forecast outcomes of matches played for the group stage. The same applies for the knock-out stage with additional information on match results prior to the round considered. For instance, in the case of the round of 16, data on the group stage matches and on the eight 1[st] leg matches played on 18-19 and 25-26 February 2013 are taken into account for predicting the eight forthcoming 2[nd] leg matches played on 11-12 and 13-19 March. A typical example of such predictions is shown on Table 4 for the match Manchester City (MCI) against Barcelona (BAR) illustrating the differences between the forecasting probabilities obtained with zero adjustment (all teams being equal in expectation) and the other two systems with a better performance for FCWR over UEFA and Zero adjustment. Notice also the difference between the posterior mean and plug-in versions of the Brier score, the former being larger due to taking into account additional uncertainty in the parameters of the distribution of $X_{ij}^{new} = k \mid \mathbf{y}^{av}$.

Table 4: An example of match forecast for the 2013-14 UEFA season: Round of 16 MCI vs BAR (0-2) 1[st] leg and BAR vs MCI (2-1) 2[nd] leg

| | Adjustment | Brier score | | Probability* | | |
|---|---|---|---|---|---|---|
| | | Post-exp | Plug-in | [1] | [X] | [2] |
| MCI-BAR | Zero | 0.932 | 0.867 | 0.527 | 0.210 | <u>0.263</u> |
| | UEFACR | 0.623 | 0.583 | 0.378 | 0.240 | <u>0.382</u> |
| | FCWR | 0.376 | 0.354 | 0.253 | 0.233 | <u>0.514</u> |
| BAR-MCI | Zero | 0.403 | 0.343 | <u>0.524</u> | 0.204 | 0.272 |
| | UEFACR | 0.199 | 0.166 | <u>0.668</u> | 0.178 | 0.154 |
| | FCWR | 0.120 | 0.109 | <u>0.732</u> | 0.161 | 0.107 |

Revised 12-02-15

[1],[X],[2] for Win, Draw and Loss respectively; Probability underlined: Accuracy or Exact Forecasting Rate ;* Posterior Predictive Probability defined as $\mathrm{E}\left[\Pr\left(X_{ij}^{new}=k\right)|\mathbf{y}^{av}\right]$

Now, we can look for the comparative efficiency of forecasting outcomes of matches for the different stages of the competition and globally using either zero pre-adjustment, or UEFACR or FCWR (table 5).

On the whole, the two rating systems provide a substantial improvement both in the Brier score (BS) and the accuracy (A) with a slightly better performance for FCWR (BS=0.530; A=47.4%) over UEFACR (BS=0.595; A=43.3%) as compared to the zero adjustment (BS=0.685; A=38.3%). As indicated on Table 5, this advantage occurs early at the group stage and round of 16, but practically, vanishes later from the quarter finals to the final when the number of matches taken into account in the prediction increases. This makes sense and reflects how the relative weights contributed by the prior and likelihood information in the forecasting evolves with time.

Table 5. Effect of the type of adjustment for the team strength upon the efficiency of predicted outcomes of CL matches (2013-2014 season)

| Phase | Type of adjustment | | | | | | | |
|---|---|---|---|---|---|---|---|---|
| | Zero | | UEFACR | | | FCWR | | |
| | Brier | Accuracy | Brier | Accuracy | | Brier | Accuracy | |
| Group stage | 0.695 | 0.377 | 0.594 | 0.434 | +15.1% | 0.524 | 0.479 | +27.0% |
| Round of 16 | 0.637 | 0.413 | 0.531 | 0.459 | +11.1% | 0.476 | 0.512 | +24.0% |
| 1/4,1/2,1/1 | 0.667 | 0.387 | 0.675 | 0.390 | +0.8% | 0.635 | 0.387 | 0.0% |
| Group+Knockout | 0.685 | 0.383 | 0.595 | 0.433 | +13.1% | 0.530 | 0.474 | +23.8% |

Accuracy : rate of exact predicted outcomes of matches (see formula 11) and % of variation with respect to the Zero adjustment for the same phase

### *4.3 Rating and ranking list of teams*

In addition, an obvious by-product of this modeling consists of editing a rating and ranking list of the 32 competing teams based solely on the outcomes of the matches they played during the group stage and the knock-out. This can be easily obtained by considering the zero prior adjustment option, all teams being equal in expectation at the beginning of the tournament. Results shown in Table 6 highlights the gap between the top seeded teams from Pot 1 and the teams qualified through the play-offs. There are a couple of outliers eg. Porto (POR) from Pot 1 but ranked only 24. On the contrary, Atletico Madrid (AMA) and Paris St Germain (PSG) from Pot 2 ranked second and third. Dortmund (DOR) and Manchester City from Pot 3 and Napoli (NAP) from Pot 4 performed also better than expected from their rating in the UEFACR. Notice also that the two finalists are ranked first and second far ahead of the following teams in agreement with the result of the final played May 24, 2014 in Lisbon.

Table 6 : Rating of the 32 teams on their match results in the UEFA CL 2013-14

| Rank | Team | Estimation | SEP | Rank | Team | Estimation | SEP |
|---|---|---|---|---|---|---|---|
| 1  | **RMA** | 2.049  | 0.968 | 17 | *ACM* | -0.018 | 0.949 |
| 2  | AMA     | 1.964  | 0.945 | 18 | JUV   | -0.049 | 0.997 |
| 3  | PSG     | 1.226  | 0.948 | 19 | *ZSP* | -0.081 | 0.917 |
| 4  | **BAR** | 1.223  | 0.954 | 20 | SHA   | -0.235 | 1.017 |
| 5  | **BAY** | 1.037  | 0.874 | 21 | BLE   | -2.888 | 0.975 |
| 6  | **CHE** | 0.803  | 0.864 | 22 | *BAL* | -0.385 | 1.024 |
| 7  | DOR     | 0.671  | 0.923 | 23 | *AWI* | -0.429 | 1.006 |
| 8  | **MUD** | 0.657  | 0.888 | 24 | *POR* | -0.465 | 1.007 |
| 9  | NAP     | 0.601  | 1.068 | 25 | KOB   | -0.701 | 1.048 |
| 10 | MCI     | 0.556  | 0.975 | 26 | *BUC* | -0.892 | 1.023 |
| 11 | OLY     | 0.473  | 0.951 | 27 | *CEL* | -1.216 | 1.106 |
| 12 | *ARS*   | 0.434  | 0.952 | 28 | AND   | -1.282 | 1.080 |
| 13 | **BEN** | 0.402  | 1.036 | 29 | PLZ   | -1.442 | 1.166 |
| 14 | GAL     | 0.291  | 0.937 | 30 | CSK   | -1.445 | 1.165 |
| 15 | *SCH*   | 0.207  | 0.936 | 31 | *RSO* | -1.683 | 1.137 |
| 16 | AJX     | -0.002 | 1.015 | 32 | OMA   | -1.967 | 1.250 |

Bold& underlined: team from Pot 1; Italics and grey shade: team qualified through the play-off; Rating value in standardized unit (mean 0 and variance 1); SEP: Standard Error of Prediction, here standard deviation of the posterior distribution of team strength

## 5. Discussion-Conclusion

We intended to build a model as simple as possible but, at the same time, make it capable of updating previous historical information easily and consistently via the Bayesian learning rule that posteriors at the end of a season can serve as priors for the next one. This is especially convenient for forecasting the outcomes of the group stage matches by combining these priors with an update of the external ratings of teams issued from UEFA or FCWR. Using the latter, we got an accuracy of 47.4% and a Brier score of 0.530 for the whole 2013-14 season. This forecasting performance might look mediocre but, as well-known from specialists, predictions of football matches are notoriously unreliable. For instance, Forrest et al (2005) reported a Brier score of 0.633 in forecasting home-draw-away match results for 5 seasons of the English football competitions. Actually, the overall forecasting performance represents a net improvement of 23.8% in accuracy and of 22.6% in Brier's score with respect to the basic situation of no prior adjustment for team strength. We implemented the UEFA-CR and FCWR rating systems as external information on team strength, but we could also have used other database systems such as the Soccer Club World Ranking (SCWR) that gave results close to FCWR for the 2013-14 season.

The choice was made of a latent variable model with a cumulative link function based on the probit, but we could have chosen another link such as the logit (Rao and Kupper, 1967). The difference between the probit and the logit is very small after adjusting for the measurement scale (logit=1.7 probit). In addition, the choice of the probit is consistent with priors on parameters of its argument (cut-off, home effect, team strength) handled as Gaussian distributions. Another option for direct modeling of the win, draw and loss would have been the "adjacent categories logit model" with the link applied to adjacent categories rather than to cumulative ones (Agresti, 1992). This approach has been proposed in sports by Davidson (1970) as another variation of the Bradley-Terry-Luce model for taking into account ties: see Shawul and Coulom (2012) for a comparison of this model with the Glenn-David and Rao-Kupper models for chess game outcomes.

Using a Bayesian approach of these models implies that team strength is automatically treated as a random effect, the benefit of which has been clearly demonstrated by several authors (eg. Harville, 1977; Cattelan, 2012). Instead of a point estimator such as REML, we have to specify a prior distribution on the inter team variability: here we chose a lognormal distribution, but other choices might have been envisioned such as an inverse gamma on the variance or a half Cauchy on the standard deviation (Gelman, 2006). The convenience for updating this prior at each season prevails in our choice. Team strength could also have been made varying over time either by introducing a dynamic stochastic process (Glickman and Stern, 1998; Coulom, 2008; Cattelan et al., 2012) or just by updating the values of the UEFA or FCWR external rating list regularly during the season. But, in that case, one must be careful not to use the same match data twice.

Table 7 : Example of inclusion of subjective information on the outcome of a forthcoming match into the model FCWR: BAY-RMA : 0-4, 1/2F, 2[nd] leg

| Source   | Weight | [1]   | [X]   | [2]   |
|----------|--------|-------|-------|-------|
| Model    |        | *0.582* | *0.219* | *0.199* |
| Expert   |        | 0.150 | 0.250 | 0.600 |
|          | 10     | 0.450 | 0.239 | 0.311 |
| Combined | 20     | **0.326** | **0.244** | **0.430** |
|          | 50     | 0.222 | 0.235 | 0.543 |
|          | 200    | 0.156 | 0.244 | 0.600 |

The model itself can be enriched with additional explanatory variables eg. importance of match or presence or not of key players. Expert's views on outcomes of football matches might also be valuable information to take into account, especially those of odds-setters (Forres et al., 2005).

Different avenues can be taken in this respect. In a Bayesian setting, elicitation of this additional expertise information can be carried out via implicit data introduced into the model; this can be easily interpreted under conjugate forms of prior and likelihood (here Dirichlet and multinomial). For a match $m$, the contribution $l(m)$ to the log-likelihood reduces simply to:

$$l(m) = \sum_{k=1}^{K=3} a_{m,k} \log(\pi_{m,k}) \tag{12}$$

where the parameter $\pi_{m,k}$ is the probability that match $m$ has outcome $k$ as modeled in (3abc) and $a_{m,k} = w_m \pi_{m,k}^{ex} - 1$ is a coefficient involving the expert's probability $\pi_{m,k}^{ex}$ that match $m$ will have outcome $k$ and $w_m$ a weight given to this expert information for match $m$.

An example of this combined model is shown in Table 7 pertaining to the Bayern Munich vs Real Madrid 2[nd] leg match of the semifinal lost by Bayern 0-4 whereas model predictions were clearly favorable to Bayern. Imagine the expert information is clearly opposite, viz. favoring Real Madrid; if integrated into the model, this probability information can change the direction of the final forecast. More work is still needed to investigate what kind of expert information is really valuable and how it should be weighted. However this is already an encouraging perspective made possible by the synthetic Bayesian approach adopted here.

Finally, it must also be noted that our procedure can be applied equally, along the same principles, to other major football tournaments eg. to the UEFA European championship and the FIFA World Cup.

**Acknowledgements-**Thanks are expressed to Prof John James (University of Sydney) and an anonymous reviewer for their critical reading of the manuscript. The author is also grateful to Mrs Andrea Craze and Mrs Martine Leduc for their English revision of the text.

**References**

Agresti, A. (1992). Analysis of ordinal paired comparison data. *Applied Statistics*, 41: 247-297.

Barnard, J., McCulloch, R., and Meng X-L. (2000). Modeling covariance matrices in terms of standard deviations and correlations with application to shrinkage. *Statistica Sinica*, 10: 1281-1311.

Cattelan, M. (2012). Models for paired comparison data: a review with emphasis on dependent data. *Statistical Science*, 27:412-433.

Cattelan, M., Varin, C., and Firth, D. (2012). Dynamic Bradley-Terry modeling of sports tournaments. *Applied Statistics*, 61: 135-150.

Coulom, R. (2008). Whole-history rating: a Bayesian rating system for players of time varying strength. *Conference on Computers and Games*, Beijing, China.

Revised 12-02-15

Davidson, R.R. (1970). On extending the Bradley-Terry model to accommodate ties in paired comparison experiments. *Journal of the American Statistical Association*, 65:317-328.

Foulley, J.L., and Jaffrezic, F. (2010). Modelling and estimating heterogenous variances in threshold models for ordinal discrete data via Winbugs/Openbugs. *Computer Methods and Programs in Biomedicine*, 97: 19-27.

Forrest, D., Goddard, J., and Simmons, R. (2005). Odds-setters as forecasters: the case of English football. *International Journal of Forecasting*, 21: 551-564.

Gelman, A. (2006). Prior distributions for variance parameters in hierarchical models. *Bayesian Analysis*, 3: 515-533.

Gelman, A., Carlin, J.B., Stern, H.S., and Rubin, D.B. (2004). *Bayesian Data Analysis*. Second Edition. London: CRC Press/Chapman and Hall.

Glenn, W.A., and David, H.A. (1960). Ties in paired-comparison experiments using a modified Thurstone-Mosteller model. *Biometrics*, 16: 86-109.

Glickman, M.E. (1999). Parameter estimation in large dynamic paired comparison experiments. *Applied Statistics*: 48, 377–394.

Glickman, M.E., and Stern, H.S. (1998). A state space model for national football league scores. *Journal of the American Statistical Association*, 93: 25-35.

Goddard, J.A. (2005). Regression models for forecasting goals and match results in association football. *International Journal of forecasting*, 21: 331-340.

Harville, D. (1977). The use of linear-model methodology to rate high school or college football teams. *Journal of the American Statistical Association*, 72: 278-289.

Laud, P.W., and Ibrahim, J.G. (1995). Predictive model selection. *Journal of the Royal Statistical Association*, 57: 247-262.

Lunn, D., Jackson, C., Best, N., Thomas, A., and Spiegelhalter, D. (2012). *The Bugs book. A practical introduction to Bayesian analysis*. London: CRC Press/Chapman and Hall.

Plummer, L. (2008). Penalized loss functions for Bayesian model comparison. *Biostatistics*, 9: 523-539.

Rao, P.V., and Kupper, L.L. (1967). Ties in paired-comparison: a generalization of the Bradley-Terry model. *Journal of the American Statistical Association*, 62: 194-204.

Shawul, D., and Coulom, R. (2012). Paired comparisons with ties: modeling game outcomes in Chess. *Web Technical report*.